\documentclass[showpacs,preprintnumbers,eqsecnum]{revtex4}
\usepackage{amsmath}
\usepackage{makeidx}
\usepackage{graphicx}
\usepackage{bm}
\usepackage{hyperref}

\begin{document}

\preprint{}
\title{Analysis of the conduction heat transfer in cantilevers under steady state cryogenic conditions}
\author{S. Spanulescu$^{1}$}
\affiliation{$^{1}$Department of Physics, Hyperion University of
Bucharest, Postal code 030615, Bucharest, Romania}

\begin{abstract}
An accurate analysis of the conduction heat transfer in a cryogenic
flask is made and some useful formulae are derived. Taking into
account the temperature dependence of conductivity and tensile
strength of the supporting rods for a helium cryostat, these
formulae may provide more exact results than the the formulae based
on simpler models. This allows the design of the supporting elements
of a liquid helium cryostat with minimum cross-section (for
minimizing the heat transfer)and proper mechanical resistance. Some
examples of numerical results and tables are also presented.
\end{abstract}

\maketitle

\section{Introduction}
Although the heat transfer theory is well known for ordinary
conditions \cite{Fou} and a huge number of studies and applications
may be found, there is still a lack of experimental data and
theoretical solutions for the problem of the heat transfer in
cryogenic systems. Taking into account the continuous development of
the cryogenic techniques \cite{Leb} and their applications
especially in superconductor systems, it becomes more and more
important to minimize the heat transfer flux and hence the loss of
cryogenic liquids in such installations.

There are two processes that enable the heat transfer to a liquid
helium cryostat built as a Dewar flask :conduction and radiation.In
figure 1 we present a possible structure of the cryostat, and one
may see that the conduction process appears due to the sustaining
elements of the inner flask. In order to minimize the heat transfer
towards such an element it has to be made using a thermal insulating
material with a cross section as small as possible. Unfortunately
this reduces the mechanical resistance of the supports, especially
at cryogenic temperatures \cite{1}. Various materials have been
experimented and it seems that the austenitic steels are quite
adequate for this purpose, despite their relatively high thermal
conductivity. Since the tests for an optimal structure are expensive
due to the liquid helium evaporation that inevitably occurs, it
would be desirable to numerically estimate the heat transfer  for
various materials, shapes and dimensions, in order to avoid as much
as possible the experimental optimization of the system.

\begin{figure*}
\includegraphics[width=5in,keepaspectratio=true]{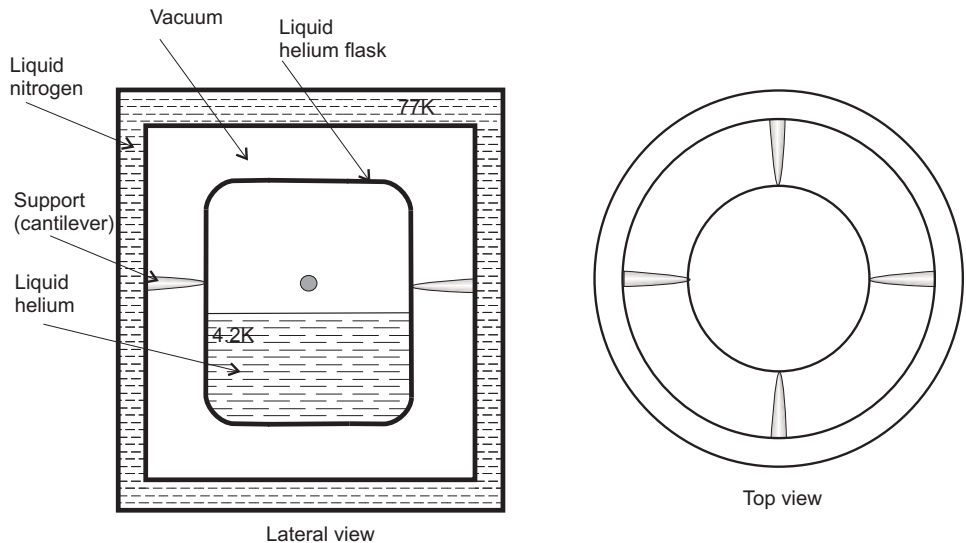}
   \caption{A typical Dewar flask for liquid helium}
   \label{figura1}
\end{figure*}

However, the usual formulae for heat transfer in rods and for their
mechanical characteristics fail in cryogenic regime, due to the
variation with the temperature of the conductance \cite{2} and
tensile strength of the materials \cite{3}. The aim of this paper is
to include as accurate as possible these variations in the formulae
necessary for the optimal design.

\vspace{-0.3cm}

\section{Heat transfer in a variable profile cantilever under cryogenic conditions}
We are interested in calculating the rate of heat transfer in a rod
placed in vacuum and with the extremities kept at two different
constant temperatures, $T_1=4.2$K  and $T_2=77$K  in steady state
conditions.

The heat transfer rate may be described by the Fourier law which
states that this is proportional to the gradient in temperature and
to the normal area through which the heat is flowing

\begin{equation}
\vec{J} =-KS\nabla T
\end{equation}
where:

$\bullet\vec{J}$  is the rate of heat transfer (heat
flux)[Wm$^{-2}$]

$\bullet K$  is the conductivity of the material
(constant)[Wm$^{-3}$K$^{-1}$]

$\bullet S$  is the cross section of the rod (constant)[m$^{2}$]

$\bullet\nabla T$ is the gradient in temperature [Km$^{-1}$].

As there is no conduction heat transfer at the lateral surface of
the rod placed in vacuum, the problem is one-dimensional and the
solution may be written as

\begin{equation}
J = \frac{{dQ}}{{dt}} = K \cdot S\frac{{(T_2  - T_1 )}}{L}
\label{12}
\end{equation}
where

$\bullet T_1$  and $T_2$ are the constant temperatures at the rod's
edges [K]

$\bullet L$ is the length of the rod [m]

This equation is valid only for constant cross section and
conductivity of the rod and small temperature difference. If this is
not the case, one must divide the whole rod in thin slices to ensure
the validity of the eq. (\ref{12}). Thus, we may write:

\begin{equation}
J = K \cdot S(x) \cdot \frac{{dT(x)}}{{dx}} \label{13}
\end{equation}

Taking also into account the dependence of the material's
conductivity with the temperature:
\begin{equation}
K = K(T(x))
\end{equation}
the equation (\ref{13} ) becomes

\begin{equation}
J = K(T(x))S(x)\frac{{dT(x)}}{{dx}}
\label{15}
\end{equation}

For cryogenic temperatures (lower than the Debye temperature), the
mean free pass of the phonons may exceed the dimensions of the rod
and hence the conductivity will be a function only of the specific
heat capacity $C_V$ \cite{2}
\begin{equation}
K(T) = \chi C_V (T)
\label{16}
\end{equation}
where the constant $\chi$ is a characteristic of the material.

According to the classical Debye model \cite{Deb}, in this regime
the specific heat capacity increases with the temperature as
\cite{2}

\begin{equation}
C_V  = aT^3
\end{equation}
involving another material's constant $a$.

More general, the Rosseland approximation \cite{Ros} \cite{Ros1}
allows considering  also an additive constant $b$ due to the
molecular conductivity

\begin{equation}
C_v  = b + aT^3
\label{17}
\end{equation}

As for $b=0$ the Debye dependence is found, we shall consider this
more general formula with the parameter $b$  more or less
significant for various materials.

From eqs. (\ref{15}), (\ref{16}) and (\ref{17}) we obtain:

\begin{equation}
J = \chi [b + aT^3 (x)]S(x)\frac{{dT(x)}}{{dx}}
\end{equation}

Separating the variables we may now integrate this ordinary
differential equation

\begin{equation}
J\frac{{dx}}{{S(x)}} = \chi [b + aT^3 (x)]dT(x)
\end{equation}

\begin{equation}
J\int\limits_0^L {\frac{{d(x)}}{{S(x)}} = \chi [b(T_2  - T_1 ) +
\frac{a}{4}} (T_2^4  - T_1^4 )] \label{18}
\end{equation}
where we took into account that the thermal flux  $J$ is constant in
steady-state conditions.

From equation (\ref{18}) we obtain the heat transfer rate $J$ as:

\begin{equation}
J = \chi [b(T_2  - T_1 ) + \frac{a}{4}(T_2^4  - T_1^4 )]\left[
{\int\limits_0^L {\frac{{dx}}{{S(x)}}} } \right]^{ - 1} \label{20}
\end{equation}

We may also obtain the temperature distribution along the rod, which
will be necessary for mechanical resistance calculations. From eq.
(\ref{15}) we may write

\begin{equation}
dT(x) = J\frac{{dx}}{{K(T(x)) \cdot S(x)}} \label{19}
\end{equation}

At a distance $l$  from the $T_1$ source, the temperature $T(l)$ is
obtained by integrating eq. (\ref{19})

\begin{equation}
\int\limits_{T_1 }^{T(l)} {\chi [b + aT^3 (x)]dT(x) =
J\int\limits_0^l {\frac{{dx}}{{S(x)}}} }
\end{equation}

\begin{equation}
\chi \{ b[T(l) - T_1 ] + \frac{a}{4}[T^4 (l) - T_1^4 ]\}  =
J\int\limits_0^l {\frac{{dx}}{{S(x)}}}
\end{equation}

Using eq. (\ref{20}) for the heat transfer rate $J$, the temperature
distribution is given by the following equation, that may be solved
numerically for obtaining the temperature distribution in the rod
\begin{equation}
bT(l) + \frac{a}{4}T^4 (l) = bT_1  + \frac{a}{4}T_1^4  + [b(T_2  -
T_1 ) + \frac{a}{4}(T_2^4  - T_1^4 )]\left[ {\int\limits_0^L
{\frac{{dx}}{{S(x)}}} } \right]^{ - 1} \int\limits_0^l
{\frac{{dx}}{{S(x)}}}
\end{equation}

\section{Considerations concerning the profile of the cantilevers in cryogenic conditions}

It is known that the general Euler-Bernoulli beam theory neglects
the share deformations and the beam deflection is given by the
equation

\begin{equation}
\frac{{\partial ^2 }}{{\partial x^2 }}(EI\frac{{\partial ^2
u}}{{\partial x^2 }}) = F
\end{equation}

where:

$\bullet E$ is the you Young modulus

$\bullet I$ is the second moment area

$\bullet F$is the distributed force (force per length)

The more general Timoshenko beam theory includes the shear forces
leading to a coupled linear partial equations:

\begin{equation}
\rho A\frac{{\partial ^2 u}}{{\partial t^2 }} = \frac{\partial
}{{\partial x}}(AkG(\frac{{\partial u}}{{\partial x}} - \theta )) +
F
\end{equation}
\begin{equation}
\rho I\frac{{\partial ^2 \theta }}{{\partial t^2 }} = \frac{\partial
}{{\partial x}}(EI\frac{{\partial \theta }}{{\partial x}}) +
AkG(\frac{{\partial u}}{{\partial x}} - \theta )
\end{equation}
where:

$\bullet \rho$  is the density of beam material;

$\bullet A$is the cross-section area;

$\bullet G$  is the shear modulus;

$\bullet k$  is the Timoshenko shear coefficient;

The tensile stress in the beam at the distance $y$ from the neutral
axis and in the same plane with the beam and the applied force is
\begin{equation}
\sigma  = Ey\frac{{\partial ^2 u}}{{\partial x^2 }} = \frac{{My}}{I}
\label{119}
\end{equation}

where $M$ is the bending moment proportional with the distance to
the applied force $F$(figure 2)
\begin{equation}
M = F(L - x) \label{120}
\end{equation}

Obviously the tensile stress must not exceed the tensile strength
$\sigma_{max}$ which is a characteristic of the material, and in
cryogenic regime depends on the temperature

Thus, from eqs. (\ref{119}) and (\ref{120}) it follows
\begin{equation}
\frac{{y_{\max } }}{I}F(L - x) \le \sigma _{\max } \label{ymax}
\end{equation}

One may see that the tensile stress increases linearly with the
distance to the applied force, being maximum at the fixed end of the
cantilever and zero at the mobile end. So, a higher mechanical
resistance is necessary towards the fixed end, meaning that the
cantilever has to be ticker as $x$ decreases. We recall that the
heat loss is increasing with the cross section of the cantilever, so
that it must be kept as low as possible.

Obviously, the solution is to use a variable cross section area
cantilever, thinner in the low tensile effort regions and thicker in
the high tensile effort ones. The ideal dependence of this cross
section should keep the same ratio between the allowed tensile
stress and the actual one in every point among the beam. This will
influence also the thermal conductivity of the cantilever and the
heat transfer flux \cite{4}.

For a variable cross section area of the beam, the second area
moment area $I$ and $y_{max}$  are also variable.

For finding the optimal profile, we must first choose the shape of
the cross-section and find the dependence of the second moment area
on its geometrical characteristics.

In most cases, the applied force is in a fixed plane and an I shape
cantilever provides the largest second moment area with a minimum
cross section area. However, in our case if the flask is inclined,
the applied force changes the relative plane and the I shape
cantilever has a much lower admissible tensile stress in such
situation.

That is why a circular or a rectangular shape of the cross section
may be a proper choice, despite its lower performance in the one
plane case.

\vspace{0.3cm}

\textbf{a) Circular shape cross section cantilever}

The second momentum area in this case is
\begin{equation}
I = \frac{{\pi r^4 }}{4}
\end{equation}
From eq. (1.21) with  $y_{\max }  = r$ we obtain
\begin{equation}
\frac{{4F}}{{\pi r^3 }}(L - x) \le \sigma _{\max } (T(x)
\end{equation}
and the profile of the cantilever is given by
\begin{equation}
r(x) \ge [\frac{{4F}}{{\pi \sigma _{\max } (T(x))}}(L -
x)]^{\frac{1}{3}}
\end{equation}

Of course, one should choose a provision for the parameter $r$ (the
radius must have a nonzero value $r_0$ at the mobile end) so that
the recommended equation is
\begin{equation}
r(x) = r_0  + [\frac{{4F}}{{\pi \sigma _{\max } (T(x))}}(L -
x)]^{\frac{1}{3}} \label{121}
\end{equation}

In figure 2 is represented the recommended dependence of the radius
on the distance to the fixed end. The most difficult part in this
equation is generated by the dependence with the temperature of the
maximum tensile strength of the material. Qualitatively, it is known
that it decreases with the temperature decrease in cryogenic regime,
but an exact formula for this dependence is not available. The
problem may be solved only numerically or using some approximation.
For numerically solving the whole problem, one must use a table
containing the values of the tensile strength of the material for
several temperature within the desired range. Now an approximating
polynomial function may be obtained with high accuracy, using for
example the method of Lagrange interpolation
\begin{equation}
\sigma _{\max } (T(x)) = \sum\limits_{i = 1}^n {\sigma _{\max } (}
T_i )\prod\limits_{\scriptstyle j = 1 \hfill \atop
  \scriptstyle j \ne i \hfill}^n {\frac{{T(x) - T_j }}{{T_i  - T_j }}}
\label{lagr}
\end{equation}

We plug in this function in eq. (\ref{121}) and together with eq
(\ref{20}) we obtain a coupled nonlinear equations that should be
solved numerically.

A simpler but less accurate solution is to approximate the
dependence of the tensile stress with a first degree polynomial in
variable $x$

\begin{equation}
\sigma _{\max } (T(x)) = \sigma _0  - cx
\end{equation}
where the constant $c$ may be obtained as the slope of a line which
best approximates the data in the mentioned table.

Thus, eq. (\ref{121}) becomes
\begin{equation}
r = r_0  + [\frac{{4F}}{{\pi (\sigma _0  - cx)}}(L -
x)]^{\frac{1}{3}}
\end{equation}

Inserting this equation in  (\ref{20}) we obtain the following
expression for the heat transfer rate in a cantilever with circular
cross section and optimal profile
\begin{equation}
J = \chi [b(T_2  - T_1 ) + \frac{a}{4}(T_2^4  - T_1^4 )] I_1
\end{equation}
where $I_1=\left[ {\int\limits_0^L {\frac{{dx}}{{S_1(x)}}} }
\right]^{ - 1}$ has an exact analytical solution that may be
expressed in terms of Gauss hypergeometric functions
$_2F_1(a,b;c;z)$ \cite{6}

\begin{eqnarray}
I_1=\frac{\sigma_0 \sqrt[3]{\frac{4 F L}{\sigma_0}+\pi  r_0} \left[4
\sqrt[3]{2} \,
_2F_1\left(\frac{1}{3},\frac{1}{3};\frac{4}{3};\frac{4 c F L+c \pi
\sigma_0 r_0}{4 c F L-4 F \sigma_0}\right)+
 \left(\frac{\sigma_0 \left(4 F+c \pi
 r_0\right)}{F (\sigma_0-c L)}\right)^{2/3}\right]}{\pi  \left(4 F+c \pi  r_0\right) \left(-\frac{4 F \sigma_0+c \pi  r_0 \sigma_0}{c F L \pi -F \pi
 \sigma_0}\right)^{2/3}}\nonumber \\
 +\frac{(c L-\sigma_0) \sqrt[3]{r_0} \left[4
 \sqrt[3]{2} \, _2F_1\left(\frac{1}{3},\frac{1}{3};\frac{4}{3};-\frac{c \pi  r_0}{4 F}\right)+\left(\frac{c \pi  r_0}{F}+4\right)^{2/3}\right]}{\left(4 F+b \pi  r_0\right) \left(\frac{c \pi
 r_0}{F}+4\right)^{2/3}}
\end{eqnarray}

\vspace{0.3cm}

\textbf{b) Rectangular shape cross section cantilever}

In the case of rectangular cross section, the second moment of area
is
\begin{equation}
I(x) = \frac{{wh(x)^3 }}{{12}}
\end{equation}
where $w$ is the width (horizontal dimension) and  $h$ is the height
(vertical dimension) of the cantilever. Since the $h$ dimension has
a greater influence to the overall resistance, only it will depend
on $x$.

Plugging in this formula in eq. (\ref{ymax}) and considering
$y_{\max } (x) = \frac{{h(x)}}{2}$, we obtain
\begin{equation}
h(x) = [\frac{{6F}}{{w\sigma _{\max } }}(L - x)]^{\frac{1}{2}}  +
h_0
\end{equation}

Again, taking into account the dependence on the temperature of the
strength of the material, we may use the Lagrange interpolation
formula (\ref{lagr}) for a polynomial approximation, or be satisfied
with the linear formula .In the last case, the profile equation
becomes

\begin{equation}
h(x) = h_0  + [\frac{{6F}}{{w(\sigma _0  - bx)}}(L -
x)]^{\frac{1}{2}}
\end{equation}

Also, inserting this equation in  (\ref{20}) we obtain the following
expression for the heat transfer rate in a cantilever with
rectangular cross section and optimal profile
\begin{equation}
J = \chi [b(T_2  - T_1 ) + \frac{a}{4}(T_2^4  - T_1^4 )] I_2
\end{equation}
where $I_2=\left[ {\int\limits_0^L {\frac{{dx}}{{wh(x)}}} }
\right]^{ - 1}$ may be numerically calculated.

\section{Numerical results and conclusions}
Using our analytical formulae for the cross sections of the
cantilever for an optimal mechanical resistance and minimal
conduction we get the numerical numerical  results in Table
\ref{cant}. We provided nonzero value at the mobile end, where the
moment is theoretically null, to prevent the shearing deformation
that was not included in the Bernoulli-Euler equation. No variation
of the tensile strength with the temperature has been considered, so
that higher values for the radius should be taken towards the mobile
end. It suggests that a conical profile would be a good
approximation for the optimum compromise between mechanical and
thermal characteristics of the cantilever.

\vspace{0.3cm}
\begin{widetext}
\begin{table*}[!ht] %
\caption{The longitudinal variation of the radius (in mm) of an austenitic steel cantilever necessary for various weights of the liquid helium flask }%
\label{cant}%
\begin{tabular}{l l l l l}
\hline\hline x[m] & \hspace{10mm} F=10N  & \hspace{10mm} F=20 N & \hspace{10mm} F=30 N & \hspace{10mm} F=40 N \\
\hline
0.01& \hspace{10mm} 3.68484  &  \hspace{10mm}4.38268  &  \hspace{10mm} 4.8722   & \hspace{10mm}5.26191\\
0.02& \hspace{10mm} 3.63688  &  \hspace{10mm}4.32226  &  \hspace{10mm} 4.80304  & \hspace{10mm}5.18579\\
0.03& \hspace{10mm} 3.58712  &  \hspace{10mm}4.25957  &  \hspace{10mm} 4.73127  & \hspace{10mm}5.10679\\
0.04& \hspace{10mm} 3.53536  &  \hspace{10mm}4.19436  &  \hspace{10mm} 4.65663  & \hspace{10mm}5.02464\\
0.05& \hspace{10mm} 3.4814   &  \hspace{10mm}4.12637  &  \hspace{10mm} 4.5788   & \hspace{10mm}4.93898\\
0.06& \hspace{10mm} 3.42499  &  \hspace{10mm}4.05529  &  \hspace{10mm} 4.49744  & \hspace{10mm}4.84943\\
0.07& \hspace{10mm} 3.36582  &  \hspace{10mm}3.98074  &  \hspace{10mm} 4.4121   & \hspace{10mm}4.7555 \\
0.08& \hspace{10mm} 3.30353  &  \hspace{10mm}3.90227  &  \hspace{10mm} 4.32226  & \hspace{10mm}4.65663\\
0.09& \hspace{10mm} 3.23768  &  \hspace{10mm}3.8193   &  \hspace{10mm} 4.22729  & \hspace{10mm}4.55209\\
0.10& \hspace{10mm} 3.1677   &  \hspace{10mm}3.73114  &  \hspace{10mm} 4.12637  & \hspace{10mm}4.44102\\
0.11& \hspace{10mm} 3.0929   &  \hspace{10mm}3.63688  &  \hspace{10mm} 4.01848  & \hspace{10mm}4.32226\\
0.12& \hspace{10mm} 3.01232  &  \hspace{10mm}3.53536  &  \hspace{10mm} 3.90227  & \hspace{10mm}4.19436\\
0.13& \hspace{10mm} 2.92471  &  \hspace{10mm}3.42499  &  \hspace{10mm} 3.77592  & \hspace{10mm}4.05529\\
0.14& \hspace{10mm} 2.82831  &  \hspace{10mm}3.30353  &  \hspace{10mm} 3.63688  & \hspace{10mm}3.90227\\
0.15& \hspace{10mm} 2.72051  &  \hspace{10mm}3.1677   &  \hspace{10mm} 3.4814   & \hspace{10mm}3.73114\\
0.16& \hspace{10mm} 2.59718  &  \hspace{10mm}3.01232  &  \hspace{10mm} 3.30353  & \hspace{10mm}3.53536\\
0.17& \hspace{10mm} 2.45113  &  \hspace{10mm}2.82831  &  \hspace{10mm} 3.0929   & \hspace{10mm}3.30353\\
0.18& \hspace{10mm} 2.26768  &  \hspace{10mm}2.59718  &  \hspace{10mm} 2.82831  & \hspace{10mm}3.01232\\
0.19& \hspace{10mm} 2.00616  &  \hspace{10mm}2.26768  &  \hspace{10mm} 2.45113  & \hspace{10mm}2.59718\\
0.20& \hspace{10mm} 2.00000  &  \hspace{10mm}2.00000  &  \hspace{10mm} 2.00000  & \hspace{10mm}2.00000\\

\hline\end{tabular}%
\end{table*}

\end{widetext}

The proposed formulae may be used for a proper design of the liquid
helium recipient in order to minimize the heat transfer and preserve
the mechanical characteristics of the system. They include the
conduction and tensile strength dependance on the temperature in
cryogenic regime which may be theoretically estimated or
experimentally determined for the material of choice. Although they
lead to a cumbersome profile of the cantilever, a good approximation
with a conical one may be considered for practical purpose, assuming
that the actual cross section is grater or equal to that predicted
by our formulae in every point.

\begin{figure*}
\includegraphics[width=3in,keepaspectratio=true]{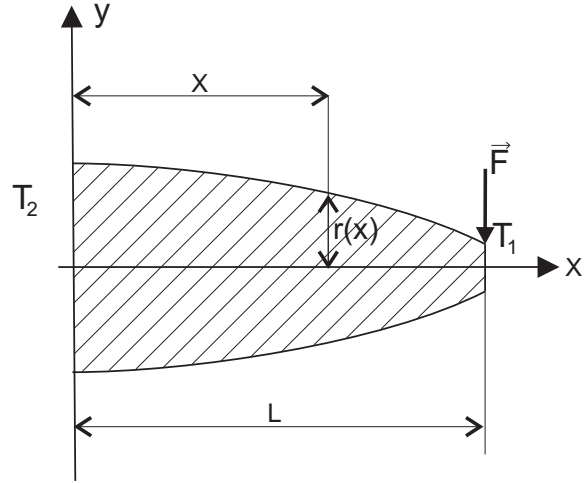}
   \caption{The optimal cantilever profile for a circular cross section}
   \label{figura2}
\end{figure*}


\acknowledgements

This work was supported by the Romanian National Research Authority
(ANCS) under Grant 22-139/2008.

%

\index{sco}

\end{document}